\documentclass[aps, pra, floatfix]{revtex4}
\usepackage{graphicx}
\usepackage{amsmath}
\usepackage{setspace}
\usepackage{color}

\begin{document}
\title{Eyring equation and the second order rate law: \\ Overcoming a paradox}

\author{L. Bonnet{\footnote{Email: claude-laurent.bonnet@u-bordeaux.fr}}
}

\affiliation{
CNRS, Institut des Sciences Mol\'eculaires, UMR 5255, 33405, Talence, France\\
Univ. Bordeaux, Institut des Sciences Mol\'eculaires, UMR 5255, 33405, Talence, France\\
} 
\date{\today}
\begin{abstract}
\noindent 
\textcolor{black}{The standard transition state theory (TST) of bimolecular reactions was recently shown to lead to a rate law in disagreement 
with the expected second order rate law [Bonnet L., Rayez J.-C. \emph{Int. J. Quantum Chem.} \textbf{2010}, 110, 2355]. An alternative 
derivation was proposed which allows to overcome this paradox. This derivation allows to get at the same time the good rate law and 
Eyring equation for the rate constant. The purpose of this paper is to provide rigorous and convincing theoretical arguments 
in order to strengthen these developments and improve their visibility.} 
\end{abstract}
\maketitle
\section{Introduction}
\label{I}

A classic result of first year chemistry courses is that for elementary bimolecular gas-phase reactions \cite{Note1} of the type 
\\
\begin{equation}
A + B \longrightarrow Products,
\label{1}
\end{equation}
\\
the time-evolution of the concentrations [A] and [B] is given by the second order rate law
\\
\begin{equation}
-\frac{d[A]}{dt} = K [A] [B],
\label{2}
\end{equation}
\\
where $K$ is the rate constant. For reactions proceeding through a barrier along the reaction path, 
Eyring famous equation \cite{Eyring}
allows the estimation of $K$ with a considerable success for more than 80 years \cite{Laidler}.
This formula is the central result of activated complex theory \cite{Eyring},   
nowadays called transition state theory (TST) \cite{Laidler,EP,Wigner1,Wigner2,Horiuti,Nakamura,Keck1,Keck2,TST1,TST2,
TST3,TST4,Mahan,Truhlar1,Truhlar2,P}. Note that other researchers than Eyring made seminal contributions to the theory at about the same time
\cite{Laidler}, like Evans and Polanyi \cite{EP}, Wigner \cite{Wigner1,Wigner2}, Horiuti \cite{Horiuti}, etc...
In the limit where the potential energy of the molecular system is separable in the vicinity of 
the barrier saddle point, Eyring equation reads 
\\
\begin{equation}
K = \frac{kT}{h}\frac{Q_{\ddagger}}{Q_A Q_B} exp\left(-\frac{U_{\ddagger}}{kT}\right),
\label{3}
\end{equation}
\\
where $k$ and $h$ are Boltzmann and Planck constants, respectively, $T$ is the temperature, 
$Q_A$ and $Q_B$ are the partition functions per unit volume of A and B, respectively, 
$Q_{\ddagger}$ is the partition function per unit volume of the activated complex or transition state (TS), 
defined as the molecular system minus the reaction coordinate at the barrier top, 
and $U_{\ddagger}$ is the classical barrier heigth with respect to the separated reagents. 
More general and sophisticated versions of Eq.~\eqref{3} are available, which include the variational search of the TS and 
quantum corrections to deal with tunneling and the quantization of both reagents and TS internal states.
The use of Eyring equation requiring electronic structure calculations for a reduced number of molecular
configurations, this formula has been applied to an uncountable number of 
widely different processes by several generations of chemists and physicists
(in comparison, molecular dynamics simulations have provided only a tiny number of rate constants, due to the heaviness of the 
numerical calculations required). In addition, the simplicity of this formula makes it an unrivaled interpretative 
tool for rationalizing the rate of chemical processes. Therefore, Eyring equation is a major result of physical-chemistry, 
taught at the graduate level for several decades in universities world-wide.

In view of this, it was a few years ago a real surprise to the authors of ref.~\cite{BR} when they realized that Eyring's derivation \cite{Eyring}, 
as well as its numerous reports in the literature since 1935, imply the following clear contradiction: while the aim of Eyring 
was to predict $K$ in Eq.~\eqref{2}, his derivation of Eq.~\eqref{3} leads to a rate law different from Eq.~\eqref{2}. This paradox was 
obviously disconcerting, but could finally be overcome by introducing in the theoretical developments a constraint on reagent species 
that was seemingly ignored at the time~\cite{BR} (see Section~\ref{III}). We wish to emphasize that the contradiction is subtle, for it does 
absolutely not put into question the validity of Eyring equation itself~\cite{BR}, as we shall see at the end of this work.

Six years later, ref.~\cite{BR} is still virtually unknown. Yet, it establishes the missing link between TST and the 
second order rate law (we could never find it in the literature), thereby shedding light on a still obscure area of reaction rate theory. 
The goal of the present work is thus essentially to strengthen the material of ref.~\cite{BR}, so it may enjoy better understanding and 
confidence among the readers concerned with the consistency of TST. 
In particular, the distinction previously made between the standard and dynamical derivations of Eyring equation~\cite{BR} is abandoned, 
for we realized that it obscures the developments and is not justified. Moreover, we shall be comparing our formal 
developments with those of authoritative works in the literature, starting with Eyring's paper, in order to check the validity of 
ours. This is a necessary work to convince the reader that the paradox raised in ref.~\cite{BR} is truly founded.

\section{Specifying the molecular system and the theoretical framework}
\label{II}

\subsection{Experimental conditions}
\label{II.A}

A vessel of volume $V$ is supposed to contain a gas made of about one mole of molecules A, the same amount of molecules B, and a much larger 
amount of inert buffer gas at a temperature $T_1$ such that the thermal energy $kT_1$ is much smaller than the barrier heigth. Hence, 
the Boltzmann population of A-B systems with an energy larger than the barrier heigth is totally negligible and reaction~\eqref{1} cannot 
take place. However, at a given instant that will be time 0 of the kinetic experiment in the next developments, 
the mixture is heated in such a way that the new 
thermal energy $kT_2$ is larger than the barrier heigth and the reaction starts. Moreover, the transient time for heating the mixture is assumed 
to be neglible as compared to the average time for reaction. The gas mixture can thus be considered to already be in canonical equilibrium 
at $T_2$ at time 0. $T_2$ will simply be called $T$ from now on. 
The newly formed products, two molecules C and D for instance, start diffusing in the gaseous medium to
eventually collide with D and C molecules, respectively. However, it is supposed that the exothermicity of reaction~\eqref{1} is sufficiently large
for $kT$ to be much smaller than the barrier for reverse reaction $C + D \longrightarrow A + B$. In other words, the equilibrium constant 
of reaction~\eqref{1} is strongly unfavorable to its reverse reaction, and the reagents are virtually not reformed. The time evolutions of 
[A] and [B] are measured and $K$ is deduced from Eq.~\eqref{2}. The electronic structure of the A-B system is such that at $T$, 
reaction~\eqref{1} takes place in the electronic ground state only.

\subsection{Potentially reactive pairs}
\label{II.B}

At time $t$ (larger than 0), the numbers of A and B molecules are $N_A$ and $N_B$, respectively. Due to thermal agitation, each A molecule 
collides a large number of times with buffer gaseous species before meeting a B molecule. This collision, however, does not necessarily 
lead to reaction. In fact, the barrier makes the reaction probability per A-B encounter rather low. A given A molecule located somewhere in the 
vessel at time 0 has thus enough time to diffuse in the gas mixture, experience many unsuccessful collisions with B species, to eventually 
react with a given B molecule located at a macroscopic distance from the initial position of A. Consequently, each A molecule can potentially 
react with each and everyone of the B molecules. The number of \emph{potentially reactive pairs} (PRPs) at time $t$ is thus 
\\
\begin{equation}
N = N_A N_B.
\label{4}
\end{equation}

\subsection{Phase space coordinates}
\label{II.C}

The A-B molecular system is assumed to involve $I$ nuclei and thus, 3$I$ Cartesian coordinates to locate them in the laboratory space. 
However, within the neigborhood of the barrier top, we shall use the normal mode coordinates at the saddle point, denoted $q_i$, $i=1,...,3I$. 
$q_1$ is the reaction coordinate associated with the imaginary frequency. $q_i$, $i=2,...,3I-6$ are vibrational coordinates.
$q_i$, $i=3I-5,...,3I$ are translational and rotational coordinates of the whole A-B system. Nuclear motions are assumed to be 
satisfyingly described by classical mechanics. This is the first assumption of TST. The momenta conjugate to $q_i$ is denoted $p_i$. 
The phase space is made of the 6$I$ coordinates $q_i$ and $p_i$, $i=1,...,3I$. $\mathbf{\Gamma}$ is a vector in this space.

\subsection{Topology of the barrier, TS and no recrossing assumption}
\label{II.D}

For simplicity's sake, we assume that the normal mode description is valid over a broad neigborhood of the saddle point. 
In this region, the topology of the potential energy surface is thus quadratic and motion is completely separable. 
As stated in the introduction, the activated complex or TS is defined by $q_1 = 0$. Within the phase space, the TS is thus 
the (hyper) surface of the phase space orthogonal to the reaction coordinate at the barrier top. Due to the previous separability, 
recrossing of the TS is not possible close to the barrier top. In addition to that, we assume that due to the strong exothermicity 
of the reaction, the repulsive forces between the nascent products C and D in the descent down the barrier are very strong and 
prevent from any return back to the reagents. Therefore, once a trajectory crosses the TS towards the products, reaction is expected 
to be completed. This is the second assumption of TST.

\subsection{Reactive pairs}
\label{II.E}

We shall call \emph{reactive pairs} (RPs) the PRPs having crossed the TS (in the product direction, necessarily).  
$N^{\ddagger}$ will be the number of RPs at time $t$.

\subsection{Canonical distribution of the states of potentially reactive pairs}
\label{II.F}

The probability distribution $\rho\left(\mathbf{\Gamma}\right)$ of the phase space states of the $N$ PRPs 
is assumed to be canonical. This is the third assumption of TST. This distribution reads 
\\
\begin{equation}
\rho\left(\mathbf{\Gamma}\right) = \frac{exp\left(-\frac{H}{kT}\right)\Theta(-q_1)}{Z(T)}
\label{5}
\end{equation}
with
\begin{equation}
Z(T) = \int \mathbf{d\Gamma} \; exp\left(-\frac{H}{kT}\right) \Theta(-q_1).
\label{6}
\end{equation}
\\
$H$ is the classical Hamiltonian or energy of the molecular system. $\Theta(x)$ is the Heaviside function equal to 0 for $x < 0$ and 1 for $x \ge 0$. 
Hence, $\Theta(-q_1)$ makes $\rho\left(\mathbf{\Gamma}\right)$ equal to 0 on the product side of the TS. The $N$ PRPs are indeed on the reagent side 
of the TS.  On the other hand, note that right at the TS, $\rho\left(\mathbf{\Gamma}\right)$ is different from 0 given our definition of the 
Heaviside function.

\subsection{Flux of reactive pairs}
\label{II.G}

The probability flux through the TS is the amount of probability crossing the TS in the product direction per unit time.
Since the no recrossing assumption is assumed to be valid, this flux is a reaction probability flux. This quantity is given by
\\
\begin{equation}
f = \int \mathbf{d\Gamma} \; \rho\left(\mathbf{\Gamma}\right)
\frac{p_1}{m_1} \delta(q_1) \Theta(p_1).
\label{7}
\end{equation}
\\
The delta term $\delta(q_1)$ limits the integral to the TS, where we have seen that $\rho\left(\mathbf{\Gamma}\right)$ is different from 0.  
Hence, the above expression makes sense. The flux of RPs, i.e., the number of RPs (or PRPs as well) crossing the TS per unit time, is thus 
\\
\begin{equation}
\frac{dN^{\ddagger}}{dt} = N f.
\label{7a}
\end{equation}
\\
This flux can also be called rate of appearence of the RPs. The rate of disappearence of the PRPs 
being equal to the previous flux, we have 
\\
\begin{equation}
 -\frac{dN}{dt} = N f.
\label{7b}
\end{equation}

\section{Paradox in Eyring theory}
\label{III}

\subsection{Reaction rate in TST}
\label{III.A}

We are now adressing the most sensitive part of the work, i.e., the definition of the reaction rate according to Eyring.
Immediately before Eq. (1) of his paper \cite{Eyring}, Eyring writes:\\

``The procedure for calculating the rate \cite{Note2} is the following:
One first calculates the concentration of activated complexes per unit length and with momentum $p$ lying between $p$ and $p+dp$, both these 
quantities taken for the degree of freedom corresponding to decomposition. This is then multiplied by the associated velocity $p/m^*$ and summed 
for all values of momenta which correspond to passing over the barrier in the forward direction, i.e., for $p$ = 0 to $\infty$.''\\

In the present work, the degree of freedom corresponding to decomposition is $q_1$, $p = p_1$ and $m^* = m_1$. 
The mathematical formulation of the above statement is not detailed by Eyring, so we give it now. 
Setting $\mathbf{\Gamma} = \left(\mathbf{\Gamma^{\ddagger}},q_1,p_1\right)$, the concentration of PRPs with the reaction coordinate
lying between $-dq_1$ and 0 and its conjugate momentum lying between $p_1$ and $p_1+dp_1$ is
\\
\begin{equation}
C^{\ddagger} = \frac{N}{V} \left[\int \mathbf{d\Gamma^{\ddagger}} \; \rho\left(\mathbf{\Gamma^{\ddagger}},0,p_1\right)\right]dq_1dp_1.
\label{8}
\end{equation}
\\
This concentration is indeed equal to the total number $N$ of PRPs multiplied by the probability that each PRP lies within the phase space 
volume defined by the reaction coordinate between $-dq_1$ and 0 and its conjugate momentum between $p_1$ and $p_1+dp_1$, all divided by the 
volume $V$ of the vessel. 
The PRPs in the previous phase space volume, stuck at the TS, are on the way to the products, and therefore, they will cross the TS and 
become reactive pairs after an infinitesimal period of time. This is the 
reason why we have used the subscript $\ddagger$ for the concentration on the left side of the above equation. 
The concentration $\frac{dC^{\ddagger}}{dq_1}$ of activated complexes per unit length 
and with momentum $p_1$ lying between $p_1$ and $p_1+dp_1$ thus reads
\\
\begin{equation}
\frac{dC^{\ddagger}}{dq_1} = \frac{N}{V} \left[\int \mathbf{d\Gamma^{\ddagger}} \; \rho\left(\mathbf{\Gamma^{\ddagger}},0,p_1\right)\right]dp_1.
\label{9}
\end{equation}
\\
Following Eyring's previous statement, 
the rate is then obtained by multiplying $\frac{dC^{\ddagger}}{dq_1}$ by $p_1/m_1$ and integrating over the positive values of $p_1$:
\\
\begin{equation}
rate = \frac{N}{V} \int \mathbf{d\Gamma^{\ddagger}}dp_1 \; \rho\left(\mathbf{\Gamma^{\ddagger}},0,p_1\right)
\frac{p_1}{m_1} \Theta(p_1).
\label{10}
\end{equation}
\\
This expression can be rewritten as
\\
\begin{equation}
rate = \frac{N}{V} \int \mathbf{d\Gamma} \; \rho\left(\mathbf{\Gamma}\right)
\frac{p_1}{m_1} \delta(q_1) \Theta(p_1).
\label{11}
\end{equation}
\\
From Eqs.~\eqref{7}-\eqref{7b} and~\eqref{11}, we arrive at:
\\
\begin{equation}
rate = \frac{d[N^{\ddagger}]}{dt} = -\frac{d[N]}{dt}.
\label{12}
\end{equation}

\subsection{Rate constant}
\label{III.B}

The rate is also known to satisfy
\begin{equation}
rate = K [A] [B].
\label{13}
\end{equation}
\\
Therefore, one deduces from Eqs.~\eqref{12} and~\eqref{13} that
\\
\begin{equation}
K = \frac{\frac{d[N^{\ddagger}]}{dt}}{[A] [B]} = -\frac{\frac{d[N]}{dt}}{[A] [B]}.
\label{14}
\end{equation}
\\
Using Eqs.~\eqref{7a} or~\eqref{7b},~\eqref{14} and the fact that
\\
\begin{equation}
[A] [B] = [N]/V,
\label{15}
\end{equation}
we arrive at
\begin{equation}
K = Vf.
\label{16}
\end{equation}
\\
From Eqs.~\eqref{5}-\eqref{7}, it may be shown after some steps of simple algebra that Eq.~\eqref{16} 
leads to Eyring famous Eq.~\eqref{3}. A luminous presentation involving these steps has been made by Mahan~\cite{Mahan}, and therefore, 
it is needless to detail them here.

\subsection{Comparison with the existing literature}
\label{III.C}

I wish to emphasize that Eqs.~\eqref{4},~\eqref{12} and~\eqref{14} are not the results of a misinterpretation of 
Eyring's developments~\cite{Eyring}. They are indeed found in the following authoritative works:\\

Reference~\cite{Keck2}: Eq.~(x) in the derivation by Keck is here denoted K.~(x) (Eq.~(x) keeps refering to the 
present work). Eq.~\eqref{4} corresponds to K.~(2.23), Eqs.~\eqref{7} and~\eqref{7b} to K.~(2.5), Eq.~\eqref{12} to 
K.~(2.6) and Eq.~\eqref{14} to K.~(2.22). In K.~(2.5), $\rho$ is in fact the canonical distribution (see Eq.~\eqref{5}) 
multiplied by $N$. In K.~(2.6), the product over final states $f$ (the products according to Keck) should be replaced by a sum. 
However, in the present work, the back reaction from $f$ to the initial state $i$ (the reagents) does not take place, so K.~(2.6) reduces 
to $dN(i)/dt = - R(f,i)$. From this equation and K.~(2.22), one gets Eq.~\eqref{14}.\\

Reference~\cite{Mahan}: Eq.~(x) in the derivation by Mahan is here denoted M.~(x). Five lines before M.~(4), Mahan 
states: ``The quantity N is the total number of A-B pairs, and so can be replaced on the right side of eqn.~(3) by $N_A N_B$, the
product of numbers of individual A and B molecules''. This is exactly the definition corresponding to Eq.~\eqref{4}. 
It is explicitely stated in the paragraph between M.~(4) and M.~(5) containing two non numbered equations that the 
rate is ``the rate of disappearance of reactant pairs expressed in terms of concentration units''. This corresponds 
to Eq.~\eqref{12}. The left equality of M.~(5) is to be compared with Eq.~\eqref{14}. \\  

Reference~\cite{Truhlar2}: Eq.~(x) in the derivation by Fern\'andez-Ramos, Miller, Klippenstein and Truhlar is here denoted FMKT.~(x).
Eq.~\eqref{4} corresponds to a synthesis of FMKT.~(2.4.19) and FMKT.~(2.4.23) (see the paragraph in between). Their left-hand-sides 
are indeed equal. Eq.~\eqref{14} corresponds to FMKT.~(2.4.22), FMKT.~(2.4.15) and FMKT.~(2.4.14) [$F(T)/V$ in FMKT.~(2.4.22) appears
to be equal to -$d[N^R]/dt$]. That the rate is given by -$d[N^R]/dt$, in agreement with Eq.~\eqref{12}, is thus implicit in their 
developments. \\

Reference~\cite{P}: Eq.~(x) in the derivation by Petersson is here denoted P.~(x). When dividing both sides of P.~(12) by $V$, 
one gets the probability 
flux $f$ (see Eq.~\eqref{16}). One thus deduces that the right-hand-side of P.~(11) is $N_A N_B f = N f$. Therefore, one recovers Eq.~\eqref{7a}. 
Remember that $f$ is the amount of probability crossing the TS per unit time, so $N f$ is the rate of disappearence of $N$. 
Consequently, one also recovers Eq.~\eqref{7b}. Eqs.~\eqref{12} and~\eqref{14} are then implicit.

These comparisons should convince the reader that Eqs.~\eqref{4},~\eqref{12} and~\eqref{14} are formal expressions 
of TST widely admitted since its birth. This clarification is important, as the paradox discussed 
in the next subsection relies on them.

\subsection{Contradiction regarding the rate law}
\label{III.D}

According to Eq.~\eqref{12}, the TST reaction rate is given by $-d[N]/dt$. However, the real rate is \emph{a priori} $-d[A]/dt$. 
Consequently, Eq.~\eqref{14} cannot lead to the second order rate law, i.e., Eq.~\eqref{2}. If one uses Eq.~\eqref{4} to replace $N$ by $N_AN_B$ 
in Eq.~\eqref{14} and the fact that 
\\
\begin{equation}
\frac{dN_A}{dt} = \frac{dN_B}{dt} 
\label{17}
\end{equation}
\\
(everytime one molecule A disappears, one molecule B disappears too), one indeed gets
\\
\begin{equation}
\frac{d[A]}{dt} = \frac{K}{V}\frac{[A] [B]}{[A]+[B]}, 
\label{18}
\end{equation}
\\
in obvious disagreement with Eq.~\eqref{2}. This is the contradiction raised in ref.~\cite{BR}.

\section{Overcoming the paradox}
\label{IV}

Suppose we do not want to use Eq.~\eqref{2} as a prerequisite in the derivation of Eyring equation~\eqref{3}, but instead, 
wish to find a derivation from first principles which provides both Eqs.~\eqref{2} and~\eqref{3} at the same time. 
In such a case, we cannot use Eq.~\eqref{13}, which derives from Eq.~\eqref{2}, and consequently, we must give up 
on the reasoning developped in Section~\ref{III}. On the other hand, we know that Eq.~\eqref{7b} is exact provided that, of course, 
the TST assumptions discussed in Section~\ref{II} are satisfied (validity of classical mechanics for the nuclei, no recrossing of the TS 
and canonical distribution of the PRP states). We thus start our new derivation from Eq.~\eqref{7b} and introduce a new ingredient 
that has been ignored in the developments until now: as a matter of fact, \emph{reactive pairs and potentially reactive pairs are correlated}. 
To understand the nature of this statement, let us number the $N_A$ molecules A and $N_B$ molecules B and 
assume that A$_1$-B$_1$ and A$_1$-B$_2$ are two potentially reactive pairs wandering about the reagent part of the phase space.
If one of these two pairs, say A$_1$-B$_1$, becomes a reactive pair, i.e., crosses the TS and reacts, A$_1$-B$_2$ ceases to be
a potentially reactive pair at the exact time of crossing for the simple reason that A$_1$ does no longer exist beyond this 
time. The same is true for A$_1$-B$_3$, ..., A$_1$-B$_{N_B}$, and also for A$_2$-B$_1$, ..., A$_{N_A}$-B$_{1}$.  Therefore, the 
total number of potentially reactive pairs lost for the reaction whenever A$_1$-B$_1$ reacts is 1 + ($N_B -1) + (N_A -1$) = 
$N_A + N_B -1$ (see Figure 1 in ref.~\cite{BR}). Since both $N_A$ and $N_B$ are of the order of 10$^{23}$, -1 can be dropped 
from the previous number. Of course, the same is true for any A-B pair crossing the TS. The rate of disappearance of the PRPs 
is thus considerably larger than expected from statistical thermodynamical arguments when the correlation between reactive and 
potentially reactive pairs is taken into account. The actual
rate of disappearance of $N$ is finally obtained by multiplying the right-hand-side of Eq.~\eqref{7b} by $(N_A + N_B)$:
\\
\begin{equation}
 -\frac{dN}{dt} = N f (N_A + N_B).
\label{19}
\end{equation}
\\
Using Eq.~\eqref{17}, we arrive at Eq.~\eqref{2} with $K$ given by Eq.~\eqref{16}, i.e., Eyring equation~\eqref{3}. In other words, 
the usual assumptions of TST and the correlation between reactive and potentially reactive pairs suffice to get at the same time the 
second order rate law and Eyring equation for the rate constant. All becomes clear.  

Note that the paradox discussed in this work will be met for any process involving a molecularity larger than 1. On the other hand, 
for unimolecular 
reactions of the type $A \longrightarrow Products$ at high pressure~\cite{TST5}, $N = N_A$ and the first order rate law is straightforwardly 
obtained from Eq.~\eqref{7b} by dividing both sides of the latter by $V$.

\section{Conclusion}
\label{V}

The transition state theory (TST) of bimolecular reactions~\cite{Eyring,Laidler,EP,Wigner1,Wigner2,Horiuti,Nakamura,Keck1,Keck2,TST1,TST2,
TST3,TST4,Mahan,Truhlar1,Truhlar2} was developped in order to predict the value of the rate constant $K$ appearing in the second order rate law
(see Eq.~\eqref{2}). This prediction is routinely made since 1935 by using the central result of TST, i.e., Eyring famous equation~\cite{Eyring} 
(see Eq.~\eqref{3}). However, the historical derivation of this expression~\cite{Eyring,EP}, reported many times in the literature for about 
80 years, is disturbing in two respects: (i) in molecular collision theory, the derivation of the second order rate law on one hand, and the 
rate constant $K$ on the other hand ($K$ is given by the Boltzmann average of the product of the reaction cross section and the reagent 
velocity vector), are inseparable and thus made at the same time~\cite{GK}. In contrast, the second order rate law is a prerequisite for 
the derivation of $K$ in TST~\cite{Eyring,Keck2,TST4,Mahan,Truhlar2}, which is unsatisfactory ; (ii) even more disturbing is the fact 
that Eyring's developpements lead to a rate law different from the second order one. However, everything becomes clear when taking into 
account, in addition to the usual assumptions of TST, what we called the \emph{correlation between reactive and potentially reactive pairs}.
Like in molecular collision theory, Eyring equation and the second order rate law are then intimately related and derived at the same time
~\cite{BR}. The contribution of this work has been to strengthen the material of ref.~\cite{BR} so as to make it more convincing. 
In particular, the distinction made in ref.~\cite{BR} between the standard and dynamical derivations of Eyring equation has been abandoned, 
for it is confusing and not justified. Moreover, we have carefully compared the developments implying the previously 
outlined contradictions with those of authoritative works in the literature, starting with Eyring's paper. Boths sets of developments 
were found to match (see Section~\ref{III}). 
We thus hope that the new derivation of Eyring equation proposed in Section 2.11 of ref.~\cite{BR} and Section~\ref{IV} of the present 
work will meet with a favourable response from the readers interested in formal aspects of reaction rate theory.



\end{document}